# Modulation of anomalous Hall angle in a magnetic topological semimetal

Jinying Yang[1,2], Yanxing Shang[1], Xingchen Liu[1,2], Yibo Wang[1,2], Xuebin Dong[1,2], Qingqi Zeng[1], Meng Lv[1], Shen Zhang[1], Yang Liu[1,2], Binbin Wang[1], Hongxiang Wei[1], Yizheng Wu[3], Stuart Parkin[4], Gangqin Liu[1], Claudia Felser[5], Enke Liu[1,*], Baogen Shen[1,6]

[1]Beijing National Laboratory for Condensed Matter Physics, Institute of Physics, Chinese Academy of Sciences, Beijing 100190, China
[2]School of Physical Sciences, University of Chinese Academy of Sciences, Beijing 100049, China
[3]Department of Physics and State Key Laboratory of Surface Physics, Fudan University, Shanghai 200433, China
[4]Max Planck Institute of Microstructure Physics, D-06120 Halle, Germany
[5]Max Planck Institute for Chemical Physics of Solids, 01187 Dresden, Germany
[6]Ningbo Institute of Materials Technology & Engineering, Chinese Academy of Sciences, Ningbo, Zhejiang 315201, China

[*]E-mail: ekliu@iphy.ac.cn

**Abstract**

**The anomalous Hall angle ($\theta^A$) is a measure of the efficiency of converting a longitudinal driving current to a transverse spin-polarized Hall current. For anomalous Hall sensing, a large anomalous Hall angle can improve the sensitivity of magnetic field detection. However, modulation of this angle is challenging and magnetic materials typically have low angles of 0.1 to 3°. Here, we report modulation of the anomalous Hall angle in the magnetic Weyl semimetal $Co_3Sn_2S_2$. We propose that the angle parameter tan$\theta^A$ can be formulated as a function of the product of electrical resistivity and anomalous Hall conductivity. Our scheme was utilized to demonstrate the modulation of tan$\theta^A$ up to a magnitude of 0.46, corresponding to an angle of around 25°. Microfabricated anomalous Hall devices using Fe-doped $Co_3Sn_2S_2$ single-crystalline nanoflakes exhibit a high Hall sensitivity of 7028 ± 341 μΩ·cm/T and a magnetic field detectability of 23.5 ± 1.7 nT/Hz$^{0.5}$ at 1 Hz.**

**Main**

## Introduction

The anomalous Hall effect (AHE) arises from an intrinsic mechanism connected to the Berry curvature of electronic bands and an extrinsic contribution from the scattering effects of impurities[1-3]. As an essential parameter of the AHE, $\tan\theta^{A} = \sigma_{H}^{A}/\sigma_{xx}$ (anomalous Hall conductivity/longitudinal conductivity) represents the conversion efficiency of the longitudinal electric current density to the anomalous Hall current density. Here, $\theta^{A}$ (the anomalous Hall angle) is the angle between the total current density ($j$) and the driving current density ($j_x$) (Fig. 1a). A large $\theta^{A}$ corresponds to a large anomalous Hall current. It is expected that materials with a large $\tan\theta^{A}$ and high spin polarization can generate large spin anomalous Hall angles[4], which makes the AHE an efficient spin current source for spin transfer torque[4,5]. In cases of anomalous Hall sensors, a large $\theta^{A}$ improves the sensitivity of the magnetic field detection[6-8].

In most magnetic materials, the magnitude of $\tan\theta^{A}$ (or $\theta^{A}$) is below 0.05 (or 3°), primarily because of the lack of the scheme and ideal materials. The $\tan\theta^{A}$ of materials used in anomalous Hall sensors is small. For example, the $\tan\theta^{A}$ of $Co_{40}Fe_{40}B_{20}$ is 0.007[7] and that of $Fe_{0.48}Pt_{0.52}$ is 0.020 [6]. Therefore, the AHE has had few technological applications to data. Compared with conventional magnetic materials, topological magnets with Weyl fermions and strong Berry curvature exhibit a large AHE and $\tan\theta^{A}$[9-12]. Some studies have reported the effects of doping on the anomalous Hall conductivity (AHC) and enhanced $\tan\theta^{A}$[13-15]. However, prior studies have typically focused on the relation between the topological band structure and the AHC. And the modulation scheme and underlying rules of $\tan\theta^{A}$ have not been investigated systematically.

In this article, we report modulation scheme of $\theta^{A}$ based on the magnetic topological material. We establish a relation between $\tan\theta^{A}$ and the easily available $\rho_{xx}$ (longitudinal resistivity) and $\sigma_{H}^{A}$, and demonstrate the modulation of $\tan\theta^{A}$ up to a magnitude of 0.46 (corresponding to $\theta^{A}$ of around 25°) in magnetic Weyl semimetal $Co_3Sn_2S_2$. The Fe-doped $Co_3Sn_2S_2$ single-crystalline nanoflakes are used to fabricate the anomalous Hall devices, which demonstrate a Hall sensitivity of 7028 ± 341 μΩ·cm/T and a magnetic field detectability of 23.5 ± 1.7 nT/Hz$^{0.5}$ at 1 Hz.

### Physical scheme proposed for tuning $\tan\theta^{A}$

Among $\rho_{H}^{A}$ (anomalous Hall resistivity), $\rho_{xx}$, $\sigma_{H}^{A}$ and $\sigma_{xx}$, any two are sufficient to determining $\tan\theta^{A}$ according to the definition and the tensor conversion under the Drude model (see details in Supplementary Section S1.1). The intrinsic and extrinsic skew scattering $\sigma_{H}^{A}$ contributions can be distinguished by the scaling relation[16,17]. The intrinsic $\sigma_{H}^{A}$ can be theoretically obtained by calculating the band Berry curvature, while the extrinsic component is related to the impurity scattering. $\rho_{xx}$ is a fundamental parameter related to various scattering and can be measured directly in electric transport. Therefore, the selection of $\sigma_{H}^{A}$ and $\rho_{xx}$ to express $\tan\theta^{A}$ has three advantages: I) $\sigma_{H}^{A}$ and $\rho_{xx}$ can be controlled by independent factors and thus modulated explicitly and separately in experiments; II) The temperature dependence of $\tan\theta^{A}$ can be predicted by theoretical calculations and resistance measurements for a given material; III) The simultaneous increase of $\sigma_{\mathrm{H}}^{\mathrm{A}}$ and $\rho_{xx}$ can be experimentally achieved in a given system. The accurate expression for $\tan\theta^{\mathrm{A}}$ is thus formulated as $\tan\theta^{\mathrm{A}} = (1+\tan^2\theta^{\mathrm{A}})\rho_{xx}\sigma_{\mathrm{H}}^{\mathrm{A}}$ (details are available in the Supplementary Section S1.1). The two solutions ($\tan\theta_1^{\mathrm{A}} = \frac{1-\sqrt{1-4(\rho_{xx}\sigma_{\mathrm{H}}^{\mathrm{A}})^2}}{2\rho_{xx}\sigma_{\mathrm{H}}^{\mathrm{A}}} \in (0,1]$ and $\tan\theta_2^{\mathrm{A}} = \frac{1+\sqrt{1-4(\rho_{xx}\sigma_{\mathrm{H}}^{\mathrm{A}})^2}}{2\rho_{xx}\sigma_{\mathrm{H}}^{\mathrm{A}}} \in [1,+\infty)$ ) of this quadratic equation show that $\tan\theta^{\mathrm{A}}$ is a function of the product ($\rho_{xx}\sigma_{\mathrm{H}}^{\mathrm{A}}$) of the longitudinal electric resistivity ($\rho_{xx}$) and anomalous Hall conductivity ($\sigma_{\mathrm{H}}^{\mathrm{A}}$). The relation between $\tan\theta^{\mathrm{A}}$ and $\rho_{xx}\sigma_{\mathrm{H}}^{\mathrm{A}}$ is plotted in Fig. 1a.

$\tan\theta^{\mathrm{A}} = 0$ is a limiting case where the anomalous Hall current is zero. Most magnetic materials with low $\tan\theta^{\mathrm{A}}$ are usually located around here. In this region, the $\tan\theta_1^{\mathrm{A}}$ increases monotonously with the increase of $\rho_{xx}\sigma_{\mathrm{H}}^{\mathrm{A}}$. The larger $\rho_{xx}\sigma_{\mathrm{H}}^{\mathrm{A}}$ is, the more the high-order terms of the Taylor expansion contribute to $\tan\theta^{\mathrm{A}}$ (Fig. S4), which makes $\tan\theta_1^{\mathrm{A}}$ increase rapidly when $\rho_{xx}\sigma_{\mathrm{H}}^{\mathrm{A}}$ exceeds 0.3. When $\rho_{xx}\sigma_{\mathrm{H}}^{\mathrm{A}}$ reaches 0.5, the two solutions meet at 1 ($\theta^{\mathrm{A}} = 45°$), where the anomalous Hall current density equals the longitudinal one. After $\tan\theta_1^{\mathrm{A}}$ exceeds 1, it continues to increase with decreasing $\rho_{xx}\sigma_{\mathrm{H}}^{\mathrm{A}}$. For the limit where $\tan\theta_2^{\mathrm{A}}$ approaches infinity, one can expect the quantum anomalous Hall (QAH) insulating state in a two-dimensional case[18]. For solution $\tan\theta_1^{\mathrm{A}}$, it dominates the metallic region where the conducting electrons play a fundamental role in electronic or thermo-electronic transport (Fig. 1a).

Our work primarily focuses on $\tan\theta_1^{\mathrm{A}}$. Figure 1b shows the three-dimensional plot of $\tan\theta^{\mathrm{A}}$ as a function of both $\rho_{xx}$ and $\sigma_{\mathrm{H}}^{\mathrm{A}}$, which exhibits a chimney-like distribution of $\tan\theta^{\mathrm{A}}$. As indicated by green and purple lines, low $\sigma_{\mathrm{H}}^{\mathrm{A}}$ results in low $\tan\theta^{\mathrm{A}}$, which corresponds to many conventional magnetic materials. Low $\rho_{xx}$ also leads to low $\tan\theta^{\mathrm{A}}$ even for materials with thousand-magnitude of $\sigma_{\mathrm{H}}^{\mathrm{A}}$ such as Fe and Ni metals[19,20]. These two cases include many materials (Figs. 1b and 3f). One can see the most efficient path to get a large $\tan\theta^{\mathrm{A}}$ is to increase $\rho_{xx}$ and $\sigma_{\mathrm{H}}^{\mathrm{A}}$ simultaneously, as shown by the red line in Fig.1b.

In order to find out practical experimental methods to achieve the large $\tan\theta^{\mathrm{A}}$, we further analyse $\rho_{xx}$ and $\sigma_{\mathrm{H}}^{\mathrm{A}}$ in real materials, as shown in Table 1. The case III is the most efficient approach for obtaining the large $\tan\theta^{\mathrm{A}}$, but it needs to be analysed specifically according to the materials. Previous works claimed that the larger $\tan\theta^{\mathrm{A}}$ could be obtained by extrapolating to the higher conductivity region using the TYJ model[16,19]. However, this method of increasing $\tan\theta^{\mathrm{A}}$ is less efficient because $\rho_{xx}$ largely decreases (case IV). For simplicity, generality and efficiency, considering that the intrinsic $\sigma_{\mathrm{H}}^{\mathrm{A}}$ and $\rho_{xx}$ can be modulated by independent variables, we analysed the necessary conditions for materials with intrinsic $\sigma_{\mathrm{H}}^{\mathrm{A}}$ (case II-1).

As an example, we consider the case of intrinsic $\sigma_{\mathrm{H}}^{\mathrm{A}}$ of 1500 $\Omega^{-1}$ cm$^{-1}$, as indicated by the orange line in Fig. 1b. The intrinsic $\sigma_{\mathrm{H}}^{\mathrm{A}}$ remains constant if $\rho_{xx}$ is increased by external variables without altering the band structure of a given system, and $\tan\theta^{\mathrm{A}}$ increases according to Figs. 1a and 1b. $\sigma_{\mathrm{H}}^{\mathrm{A}}$ decays ($\sigma_{\mathrm{H}}^{\mathrm{A}} \propto \sigma_{xx}^{1.6}$, see unified model in Fig. S7[21,22]) if $\rho_{xx}$ increases further into the hopping region, resulting in a decrease in $\tan\theta^{\mathrm{A}}$.

A maximum of $\tan\theta^{A}$ can be obtained at the transition region connecting the intrinsic and hopping regions. The relation between $\sigma_{H}^{A}$ ($\tan\theta^{A}$) and $\rho_{xx}$ can be deduced from Fig. 1b, as shown in the *xy*-plane (*yz*-plane) projection in Fig. 1b (Fig. 1c). For general understanding, additional curves with intrinsic $\sigma_{H}^{A}$ are presented in Figs. S6 and 1c.

The transition region locates in a range (for example, 150 - 350 μΩ cm), that depends on the materials and modulating approaches. For a given material system, in order to obtain a large $\tan\theta^{A}$, the intrinsic region should be as wide as possible before $\rho_{xx}$ enters the transition region. Figure 1c further shows that for larger values of the intrinsic $\sigma_{H}^{A}$, the maximum of $\tan\theta^{A}$ increases in the transition region.

**Modulation scheme with designed degrees of freedom**

The determination of the degrees of freedom is essential for the experimental realization of large $\tan\theta^{A}$. For simplicity, we start from the modulation of a single variable and give selected approaches based on the nature of the intrinsic $\sigma_{H}^{A}$ and $\rho_{xx}$. Clearly, the fastest way to increase $\tan\theta^{A}$ is to simultaneously increase both $\sigma_{H}^{A}$ and $\rho_{xx}$ (red line in Fig. 1b). As tuning approaches often have various effects on material properties, proper experimental approaches are required to achieve this simultaneous increase of $\sigma_{H}^{A}$ and $\rho_{xx}$. Therefore, different degrees of freedom that closely correlate $\rho_{xx}$ and $\sigma_{H}^{A}$ were proposed for this design (Fig. 2).

Modulation of intrinsic $\sigma_{H}^{A}$: The intrinsic $\sigma_{H}^{A}$ is determined by the electronic band structure, which is affected by the crystal field, exchange interaction, and spin-orbital coupling (SOC) of a system[23]. When the Fermi level is tuned into the SOC-opened anti-crossing gap in a topological magnet, the intrinsic $\sigma_{H}^{A}$ will be enhanced resonantly[3,9,21]. This can be experimentally achieved by gating, electron- or hole-doping[13-15,18,24].

Modulation of $\rho_{xx}$: The large Berry curvature around the Fermi level is always obtained in topological magnets[9-11,25]. Furthermore, based on the large intrinsic $\sigma_{H}^{A}$, the thickness ($t$)[16,26,27], temperature ($T$)-dependent factors (interactions of electron-electron[28], electron-phonon[29], electron-magnon[30]), doping, vacancy, or grain boundary are potential scattering manners, which alter the relaxation time ($\tau$), for modulating $\rho_{xx}$ without significantly altering the electronic structures, corresponding to case II-1 in Table 1. $\tan\theta^{A}$ will increase with the decrease (or increase) of the variables ($t$, $T$, concentrations of vacancy or grain boundary).

Modulation of both $\rho_{xx}$ and $\sigma_{H}^{A}$: It is worth strengthening that the chemical doping is a special experimental method that distinctively affects $\tan\theta^{A}$[13-15]. Doping can increase longitudinal $\rho_{xx}$ by enhancing scattering; simultaneously, it increases transverse $\sigma_{H}^{A}$ by introducing the extrinsic contribution or, even intriguingly, enhancing the intrinsic contribution owing to band broadening or Fermi level shifting[13,14]. These cases are related to case III in Table 1. In this case, $\tan\theta^{A}$ may rapidly increase, as expected from the red path in Fig. 1b.

**Realization of large $\tan\theta^{A}$ and sensing demonstration**

Based on our proposed formula and design scheme, $Co_3Sn_2S_2$ is selected for experimental validation, which is a typical magnetic Weyl semimetal with giant intrinsic $\sigma_{\mathrm{H}}^{\mathrm{A}}$[9,31]. We began with the large intrinsic $\sigma_{\mathrm{H}}^{\mathrm{A}}$ and engineered the $\rho_{xx}$ to enter the transition region through temperature-dependent scattering and reduction of the sample thickness. In this case we obtained the large intrinsic tan$\theta^{\mathrm{A}}$ of $Co_3Sn_2S_2$. Furthermore, we enhanced tan$\theta^{\mathrm{A}}$ by introducing extrinsic $\sigma_{\mathrm{H}}^{\mathrm{A}}$ and increasing $\rho_{xx}$ via Fe doping and decreasing the thickness. Figure 3 presents the experimental results of modulating tan$\theta^{\mathrm{A}}$ in $Co_3Sn_2S_2$ by different parameters (details of samples and measurements are provided in Methods and Supplementary Section S2 and S3). The temperature-dependent magnetisations for bulks and nanoflakes confirm the robust magnetic order even in the nanoflakes (Fig. 3a).

The temperature is an effective factor that can be easily applied in experiments. First, the effects of temperature on $\sigma_{\mathrm{H}}^{\mathrm{A}}$ and $\rho_{xx}$ of $Co_3Sn_2S_2$ are shown in Fig. 3b. Owing to the stable ferromagnetic state and electronic structure, the intrinsic $\sigma_{\mathrm{H}}^{\mathrm{A}}$ is independent of temperature below 90 K[32,33]. However, $\rho_{xx}$ increases significantly with the temperature due to the increase of electron-magnon scattering. Our fitting result is consistent with the results of inelastic neutron scattering[34] and electronic transport[35] in $Co_3Sn_2S_2$ reported previously (Supplementary Section S2.2.2). In Fig. 3c, tan$\theta^{\mathrm{A}}$ increases significantly with $\rho_{xx}$ from 2 to 90 K, which is consistent with Fig. 1c and corresponds to the case II-1 in Table 1. This behaviour can be also obtained in other systems (Fig. S12).

Reducing the thickness of the nanoflake samples is another method to increase $\rho_{xx}$ [16,27], which can be used to drive $\rho_{xx}$ to enter the transition region for the largest intrinsic tan$\theta^{\mathrm{A}}$. As $\rho_{xx}$ increases, tan$\theta^{\mathrm{A}}$ of $Co_3Sn_2S_2$ nanoflakes increases to nearly 0.3 by reducing the thickness and increasing the temperature (Figs. 3d and S16). This behaviour, in which tan$\theta^{\mathrm{A}}$ exhibits a maximum in the transition region, is also observed in Sb-doped $Co_3Sn_2S_2$ bulks[15] (inset in Fig. 3d), where $\sigma_{\mathrm{H}}^{\mathrm{A}}$ is primarily dominated by the intrinsic mechanism. This intrinsic approach (case II-1 and Fig. 1c) is thus effective and universal for tuning tan$\theta^{\mathrm{A}}$. Therefore, for a material with an intrinsic $\sigma_{\mathrm{H}}^{\mathrm{A}}$, the $T$- (or other parameters) dependent tan$\theta^{\mathrm{A}}$ can be predicted if the correspondence between $\rho_{xx}$ and $T$ (or other parameters) is obtained. This is advantageous for the film-device applications of $\theta^{\mathrm{A}}$, where the maximal tan$\theta^{\mathrm{A}}$ and excellent properties can be obtained around the working temperature in a suitable thickness for the materials.

As another approach to increase the product of $\sigma_{\mathrm{H}}^{\mathrm{A}}$ and $\rho_{xx}$, Fe-doped $Co_3Sn_2S_2$ nanoflakes were studied considering the special effects of chemical doping mentioned above. Owing to the increase in the doping level and the decrease in the thickness, $\rho_{xx}$ at 2 K increases to 200-300 μΩ cm (Fig. 3e), meanwhile, $\sigma_{\mathrm{H}}^{\mathrm{A}}$ remains at a high level of approximately 1300 $\Omega^{-1}$ cm$^{-1}$ even higher value up to 1600 $\Omega^{-1}$ cm$^{-1}$ (Fig. 3e). The components of $\sigma_{\mathrm{H}}^{\mathrm{A}}$ cannot be separated by TYJ model[16] because $\rho_{xx}$ enters the transition region (Fig. S18d) and the intrinsic $\sigma_{\mathrm{H}}^{\mathrm{A}}$ is not constant anymore. By comparing the curves of tan$\theta^{\mathrm{A}}$-$\rho_{xx}$ of Fe-doped $Co_3Sn_2S_2$ nanoflakes with that of intrinsic theoretical trend of pristine $Co_3Sn_2S_2$, it was found that the extrinsic contribution was introduced by impurity scattering, which is consistent with the case of Fe-doped $Co_3Sn_2S_2$ bulks (see more details in Figs. S17 and S18). Owing to the simultaneous increase in $\rho_{xx}$ and $\sigma_H^A$, $\tan\theta^A$ increases rapidly, as expected from Figs. 1a and 1b, blue line in Fig. 3e and case III in Table 1. The measured values are improved by a considerable margin (Figs. 3e and 3f). A $\tan\theta^A$ up to 0.46, corresponding to $\theta^A \approx 25°$, is an order of magnitude higher than those of many magnetic materials. The data of zero-field $\tan\theta^A$ from the literature and our results are presented in Fig. 3f (see details in Supplementary Section S4). One can see $\tan\theta^A$ of most conventional materials are below 0.05, which indicates that they are located around the corner of $\rho_{xx}$-$\sigma_H^A$ plane (purple and blue curves), leaving a blank region above the curve. With the aid of the scheme, clearly, our results successfully enter this blank region with a high value of 0.46.

In the past seventy years, the situation of low $\theta^A$ remains unchanged after the first report of anomalous Hall angle[36]. After the first enhancement in $\theta^A$ at the emergence of magnetic Weyl semimetal (red pentagram), our proposal of scheme led to the second and giant jump (red-yellow pentagrams in Fig. 4a), which enables $\theta^A$ of magnetic metals to climb to a brand-new level. According to the voltage sensitivity formula of anomalous Hall sensors (see Eq. S9), a large $\theta^A$ can generate remarkable voltage change under the same conditions of longitudinal voltage, saturation field, and dimensional ratio of devices, which can intrinsically promote the signal-to-noise performance in key chips against the additional operational amplifiers usually used in sensing circuits. In order to validate the large anomalous Hall angle effect, we fabricated anomalous Hall sensors using Fe-doped $Co_3Sn_2S_2$ single-crystalline nanoflakes. The anomalous Hall voltage is utilized to detect the external magnetic field. The magnetic field detectability ($23.5 \pm 1.7$ nT/Hz$^{0.5}$ at 1 Hz) was obtained owing to high sensitivity ($7028 \pm 341$ μΩ·cm/T) and low Hall voltage noise (Figs. 4b and S21), exhibiting a competitive advantage compared with the materials used in current anomalous Hall sensors (Fig. 4c). The results demonstrate a great potential for sensing applications of giant anomalous Hall angle (inset of Fig. 4c).

## Conclusions

We have proposed an expression of $\tan\theta^A$ with $\rho_{xx}$ and $\sigma_H^A$ and reported a modulation scheme and a picture of the anomalous Hall angle. Based on this, several design approaches were demonstrated in the magnetic Weyl semimetal $Co_3Sn_2S_2$. We obtained a $\tan\theta^A$ up to a magnitude of 0.46, corresponding to a $\theta^A$ of around 25°. We also fabricated anomalous Hall sensing devices and demonstrated a sensitivity of 7028 $\pm$ 341 μΩ·cm/T and a magnetic field detectability of $23.5 \pm 1.7$ nT/Hz$^{0.5}$ at 1 Hz. Our scheme could potentially be used to modulate giant anomalous Hall currents and even giant spin Hall currents[37,38] in other topological magnets. The scheme of giant anomalous Hall angle proposed in this study will promote the advanced applications of emerging topological materials in magneto-electronics and spintronics[39-41], such as automatic intelligent vehicles and hard disk readers[8].

## Methods

### Crystal growth and device fabrication

The bulk single crystals of $Co_3Sn_2S_2$ were grown by self-flux methods that the Sn was flux. High purity Co (99.98%), Sn (99.999%) and S (99.99%) were put into alumina crucible and sealed into a high vacuum quartz tube to prevent oxidation at high temperature. The samples were heated to 1050 °C and kept for 24 hours, and then cooling to 700 °C for centrifugation. The compositions of $Co_3Sn_2S_2$ single crystals were checked by the energy dispersive x-ray spectroscopy (EDS). The crystal structure was characterized by x-ray diffraction (XRD) at room temperature, which indicates the grown crystals have good quality (Fig. S8). The structural analysis indicates the space group of *R*-3*m* (no. 166) for $Co_3Sn_2S_2$, in which the $Co_3Sn$ forms the quasi-2D lattice, which forms a sandwich structure with S atoms.

The pristine and Fe-doped $Co_3Sn_2S_2$ single-crystalline nanoflakes were grown on $Al_2O_3$ substrates by chemical vapor transport (CVT) method. The polycrystalline raw material and transport agent were sealed into the quartz tube in a high vacuum to grow single crystal nanoflakes under different temperature gradients. Hexagonal flakes with diameters of 10 ~ 100 μm and thicknesses of 10 ~ 100 nm were obtained. The atomic force microscopy (AFM) was used to characterize the thickness of nanoflakes. The $Co_3Sn_2S_2$ nanoflakes have good regular morphology (Fig. S13), which makes it easy to fabricate the hall bar pattern along *a*-axis on the sample (Fig. S9). Electron beam lithography and ion beam etching were used to fabricate the Hall bar geometry devices. Electron beam evaporation was used to evaporate the electrodes (Ti/Au).

### Magnetisation measurements

The Quantum Design vibrating sample magnetometer (VSM) option based on Physical Property Measurement System (PPMS) was used to measure the magnetisation of $Co_3Sn_2S_2$ bulks. Standard samples were prepared for magnetic measurements. The temperature dependent magnetisation from 2-300 K with the magnetic field direction along the *c*-axis is shown in Fig. 3a.

The magnetisation of $Co_3Sn_2S_2$ nanoflake was measured by the diamond nitrogen vacancy (NV) centers. Nanodiamonds (NDs) with ensemble NV centers (Adámas Nanotechnologies, NV concentration ~ 3 ppm, ND diameter ~ 100 nm) were dropped onto the surface of the $Co_3Sn_2S_2$ nanoflakes (Fig. S15c). The sample was then loaded into a closed-cycle optical cryostat (Montana Instruments, S100) for low-temperature measurements. Optically detected magnetic resonance (ODMR) spectra were obtained for each selected NDs by recording their NV fluorescence while simultaneously scanning the frequency of the applied microwave signal. These spectra were fitted with a multi-peak Lorentz function and the splitting between the outermost resonant dips was used to estimate the magnetic field strength, as shown in Figs. S15d and 3a.

### Transport measurements

The $Co_3Sn_2S_2$ single crystal samples were prepared in long strips for electronic transport measurements, which were performed on PPMS resistivity option. The transport measurements of nanoflakes were performed with Quantum Design Physical Properties Measurement System (PPMS), Keithley 2400 and Keithley 2182A. The linear $I$-$V$ curves were confirmed before measurements. In all electrical transport measurements, the current is along the $x$- ($a$-) axis ($[2\bar{1}\bar{1}0]$), the Hall measurements are along the $y$-axis ($[01\bar{1}0]$), and the magnetic field is parallel to the $z$- ($c$-) axis ([0001]) (Fig. S9). All the measurement results of longitudinal (transverse) were symmetrized (anti-symmetrized).

**Anomalous Hall sensing measurements**

The low-temperature environment for anomalous Hall sensing measurements is provided by PPMS. The alternating-current with the effective value of 1 mA was applied in the $x$ direction of the sample as the driving current, which was generated by the lock-in amplifier (Zurich Instruments, MFLI). The transverse ($y$ axis) anomalous Hall voltage was measured by the lock-in amplifier ($x$ and $y$ axes are shown in Fig. S20). Then the periodogram was employed to estimate the power spectral density of anomalous Hall voltage ($S_V$). Repeated anomalous Hall voltage measurements were performed to obtain the expectation of $S_V$. The magnetic field detectability $S_T^{0.5}$ can be calculated by $S_T^{0.5} = S_V^{0.5}/Is$, $I$ is the driving current, $s=\Delta R^A/\Delta(\mu_0 H)$, and $R^A$ is the anomalous Hall resistance.

**Data availability**

The data that support the findings of this study are available from the corresponding authors upon reasonable request. Source data are provided with this paper.

**Acknowledgements**

We thank Binghai Yan for valuable discussions. This work was supported by the State Key Development Program for Basic Research of China (No. 2022YFA1403800, to B.W.), the Fundamental Science Center of the National Natural Science Foundation of China (No. 52088101, to E.L.), the CAS Project for Young Scientists in Basic Research YSBR-057 (to M.L.), the National Natural Science Foundation of China (No. 12174426, to H.W.), the Strategic Priority Research Program (B) of the Chinese Academy of Sciences (CAS) (No. XDB33000000, to E.L.), the Synergetic Extreme Condition User Facility (SECUF, to E.L.), and the Scientific Instrument Developing Project of CAS (No. ZDKYYQ20210003, to E.L.).

**Author contributions**

E.L. conceived and supervised the project. X.D. and M.L. synthesized $Co_3Sn_2S_2$ bulks. J.Y., Q.Z. and S.Z. grown the nanoflakes. J.Y., Y.Wang and Y.L. carried out XRD measurement. J.Y., Y.Wang performed the magnetic measurements of bulks by VSM. Y.S. and G.L. performed the magnetic measurements of nanoflakes by NV centers. J.Y. and X.L. performed the fabrications of Hall devices and the measurements of anomalous Hall sensing. J.Y., B.W., H.W., Y.Wu, S.P., G.L., C.F., E.L. and B.S. analysed and discussed the experimental data. J.Y. and E.L. wrote this manuscript with the input from all the authors.

**Competing interests**

The authors have applied for a Chinese patent (no. 2025101128621) for anomalous Hall sensors based on large anomalous Hall angle and their preparation methods, which is now under consideration.

**Table 1. Different cases for tuning tan$\theta^A$ as function of product of $\sigma_H^A\rho_{xx}$.** "↑", "↓" and "→" represent an increase, decrease and invariability of parameters, respectively. More arrows mean faster increase.

| Case | $\sigma_H^A$ | $\rho_{xx}$ | $\sigma_H^A\rho_{xx}$ | tan$\theta^A$ | Approaches |
|---|---|---|---|---|---|
| I | ↓ | ↑ | ↑ | ↑ | |
| II-1 | → | ↑ | ↑ | ↑↑ | $Co_3Sn_2S_2$ (Intrinsic, temperature ($T$) and thickness ($t$)) |
| II-2 | ↑ | → | ↑ | ↑↑ | |
| III | ↑ | ↑ | ↑ | ↑↑↑ | Fe-doped $Co_3Sn_2S_2$ (Intrinsic + Extrinsic, doping) |
| IV | ↑ | ↓ | ↑ | ↑ | Extrapolation using TYJ model[16] (Extrinsic + Intrinsic) |

## Figure Legends/Captions

**Fig. 1. Generic formula and scheme for anomalous Hall angle ($\theta^{A}$). a**, Relation between tan$\theta^{A}$, longitudinal resistivity ($\rho_{xx}$) and anomalous Hall conductivity ($\sigma_{H}^{A}$). Absolute values of tan$\theta^{A}$ and $\sigma_{H}^{A}$ are adopted here. Two solutions ($\theta_{1}^{A}$ and $\theta_{2}^{A}$) connect the two regions and the split line is tan$\theta^{A}$ = 1. For most materials, tan$\theta^{A}$ is usually smaller than 0.05. The dotted lines indicate the first ($f^{1}$), third ($f^{3}$), fifth ($f^{5}$)-order terms of the Taylor expansion. $j_{y}^{A}$ and $j_{x}$ are anomalous Hall current density and longitudinal current density, respectively. The shading, which changes from white to blue to red, roughly indicates an increase in tan$\theta^{A}$. **b**, Three-dimensional plot of tan$\theta^{A}$ as a function of $\rho_{xx}$ and $\sigma_{H}^{A}$. The green (purple) curve represents tan$\theta^{A}$ as a function of $\rho_{xx}$ ($\sigma_{H}^{A}$) when $\sigma_{H}^{A}$ = 180 $\Omega^{-1}$ cm$^{-1}$ ($\rho_{xx}$ = 30 μΩ cm). In both cases, tan$\theta^{A}$ increases slowly with increasing variable when $\sigma_{H}^{A}$ and $\rho_{xx}$ are low. In contrast, tan$\theta^{A}$ rapidly increases (red curve) when $\rho_{xx}$ and $\sigma_{H}^{A}$ increase simultaneously. The tan$\theta^{A}$ increases with $\rho_{xx}$ along the line of intrinsic $\sigma_{H}^{A}$ = 1500 $\Omega^{-1}$ cm$^{-1}$ and then decreases owing to the decay of the intrinsic component of $\sigma_{H}^{A}$ during parameter tuning (orange curve). The *xy*-plane projection of the orange curve demonstrates a clear dependence of $\sigma_{H}^{A}$ on $\rho_{xx}$. **c**, Relation between tan$\theta^{A}$ and $\rho_{xx}$ extracted from the *yz*-plane projection of **b**. The dashed lines in the middle correspond to the transition region from the intrinsic to the hopping region. The orange line with intrinsic $\sigma_{H}^{A}$ = 1500 $\Omega^{-1}$ cm$^{-1}$ is the same as that in **b**. 0.5, 1.0, 1.5 and 2.0 are the values of intrinsic anomalous Hall conductivity $\sigma_{H}^{in}$ ($10^{3}$ $\Omega^{-1}$ cm$^{-1}$) in the intrinsic region.

**Fig. 2. Proposed degrees of freedom for tuning $\rho_{xx}\sigma_{\mathrm{H}}^{\mathrm{A}}$ based on topological magnets**. As tan$\theta^{\mathrm{A}}$ ($\theta^{\mathrm{A}}$ is anomalous Hall angle) is a physical quantity that is related to both the intrinsic and extrinsic factors, we analyse the influence of modulation methods on tan$\theta^{\mathrm{A}}$ from the perspective of intrinsic band structure and extrinsic scattering. Anomalous Hall conductivity ($\sigma_{\mathrm{H}}^{\mathrm{A}}$) can be modulated by doping and gating, which affects the electronic structure and associated Berry curvature ($\Omega$). We can even grasp longitudinal resistivity ($\rho_{xx}$) as the single variable to tune the tan$\theta^{\mathrm{A}}$ if using materials with giant intrinsic $\sigma_{\mathrm{H}}^{\mathrm{A}}$ instead of directly tuning the product variable $\rho_{xx}\sigma_{\mathrm{H}}^{\mathrm{A}}$. The methods for decreasing the relaxation time ($\tau$) and increasing $\rho_{xx}$ include temperature dependent scattering, doping, reduced thickness, vacancy, and grain boundary.

**Fig. 3. Experimental modulations of anomalous Hall angle ($\theta^{A}$) in $Co_3Sn_2S_2$. a**, Temperature ($T$) dependence of the magnetization ($M$) of bulk at 0.4 T measured by vibrating sample magnetometer (VSM). Local magnetic field ($\mu_0 H$) detected by the NV sensors, which is proportional to the stray field of the nanoflake sample with a thickness of 17 nm in zero field. The bulks and flakes have consistent Curie temperature ($T_C$) at 175 K. **b**, Temperature dependences of longitudinal resistivity ($\rho_{xx}$) and anomalous Hall conductivity ($\sigma_H^A$) of $Co_3Sn_2S_2$ bulks at 0.4 T. **c**, The increase in tan$\boldsymbol{\theta}^{\mathbf{A}}$ induced by the increase in $\rho_{xx}$ as the temperature increases up to 90 K at 0.4 T. **d**, tan$\theta^{A}$ of $Co_3Sn_2S_2$ nanoflakes with different thicknesses measured at 0.1 T below 90 K (pink: 81.7 nm; cyan: 17.4 nm; orange: 14.2 nm). Inset shows the $\rho_{xx}$-dependent tan$\theta^{A}$ of $Co_3Sn_{2-y}Sb_yS_2$ bulks (Black, red, blue, green and purple curves correspond to $y$=0, 0.1, 0.3, 0.5, 0.7, respectively)[15]. **e**, Enhanced $\rho_{xx}$ and $\sigma_H^A$ by decreasing the thickness and Fe doping (green: Pristine flake with a thickness of 81.7 nm; red: Fe-doped flakes). Inset shows the $\sigma_H^A \rho_{xx}$ dependence of tan$\theta^{A}$ of $Co_3Sn_2S_2$ nanoflake and Fe-doped $Co_3Sn_2S_2$ nanoflakes. The Hall loop, thickness, level of Fe doping are shown in Fig. S18. The grey dotted line is $\sigma_H^A$ of pristine $Co_3Sn_2S_2$ flake with a thickness of 81.7 nm. The blue line in the inset is the curve of $\theta_I^A$ in Fig. 1a. **f**, Enhancement of tan$\theta^{A}$ at 2 K by doping and surface scattering (Red pentagram stars denote Fe-doped nanoflakes). Comparison of tan$\theta^{A}$ from previous reports and our results. The red triangles represent the pristine $Co_3Sn_2S_2$[9] bulk. Contour lines in the $\rho_{xx}$-$\sigma_H^A$ plane represent tan$\theta^{A}$. The information about the materials is presented in Table S1.

**Fig. 4. Zero-field anomalous Hall angles ($\theta^A$) and anomalous Hall sensing**. **a**, Development of the zero-field $\theta^A$ of our results ($Co_{3-x}Fe_xSn_2S_2$ nanoflakes) and those of conventional metals. The data of all the materials are shown in Table S1. The purple and gray arrows indicate the direction of the current density ($j$). The red and blue arrows indicate the direction of the electron movement and the direction of the longitudinal electric field ($E_x$), respectively. The shading means increase in $\theta^A$ with year. **b**, Magnetic field detectability ($S_T^{0.5}$) spectra of $Co_{2.98}Fe_{0.02}Sn_2S_2$ nanoflake at 173.3 K. Inset shows the Hall device with the white scale bar of 20 μm. Inset is the transverse resistance and the purple/blue dots are the locations where the noise spectra were measured (see details in Methods and supplementary Section S5). The purple/blue curves correspond to the purple/blue dots in the inset. **c**, Distribution of magnetic field detectability (at 1 Hz) and anomalous Hall resistivity sensitivity. The purple and blue stars correspond to the purple and blue curves in b. In order to exclude the thickness factor of devices, the $\Delta\rho_H^A/\Delta(\mu_0 H)$ was used to represent the anomalous Hall resistivity sensitivity. The gray and black error bars (standard deviation) are the error bars of $S_T^{0.5}$ and anomalous Hall resistivity sensitivity, respectively. The $S_T^{0.5}$ are 23.5 ± 1.7 nT/Hz$^{0.5}$ (purple) and 57.0 ± 5.4 nT/Hz$^{0.5}$ (blue), respectively. The error of $S_T^{0.5}$, calculated using the error propagation formula, arises from the repeated Hall voltage measurements and the linear fitting of the field-dependent anomalous Hall resistance. Even near the Curie temperature of nanoflakes, the advantage of materials with large $\theta^A$ in magnetic field detection is demonstrated. Inset shows the schematic of AHE sensing reader.

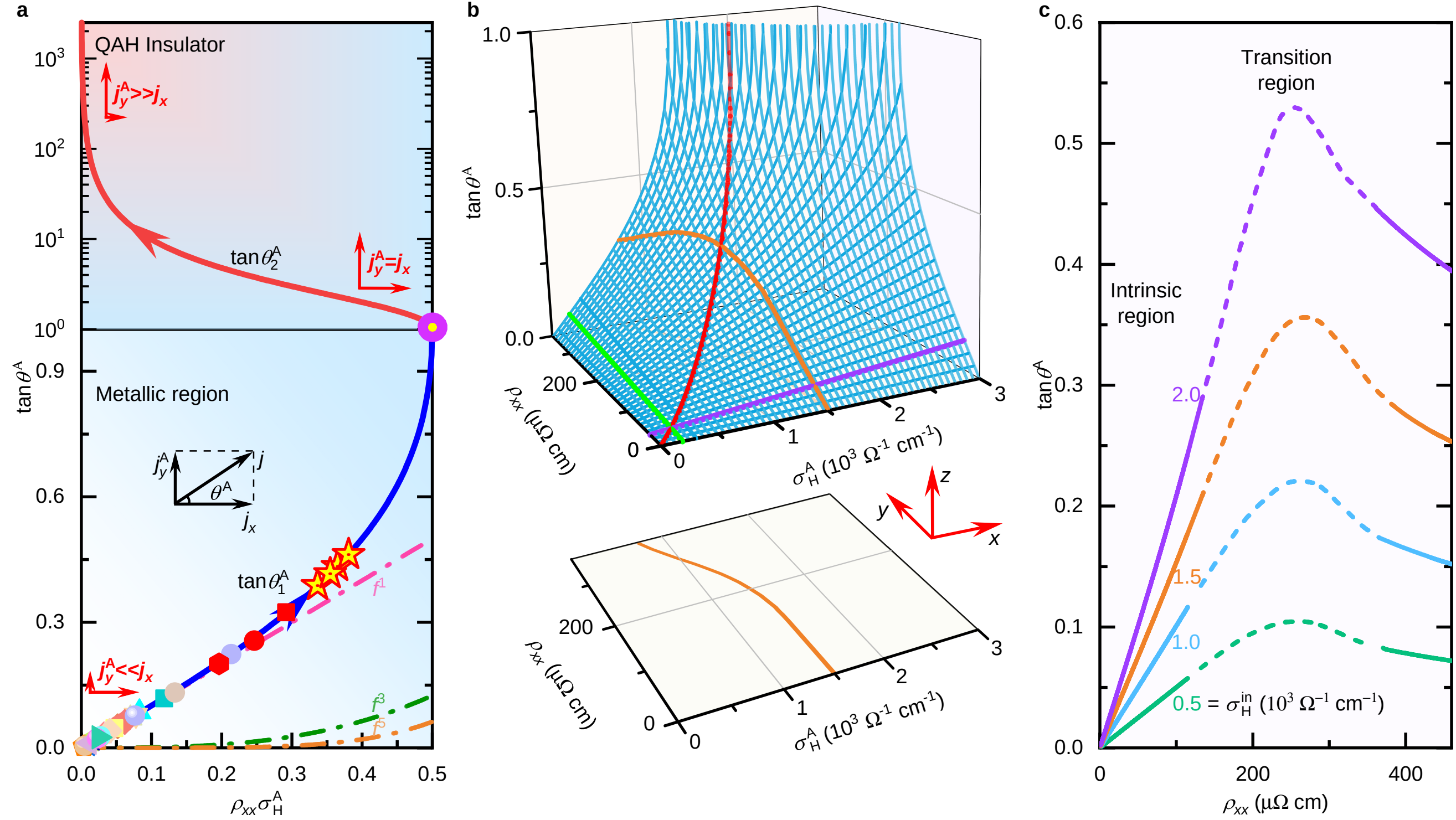

a
QAH Insulator
$j_y^A >> j_x$
$\tan\theta_2^A$
$j_y^A = j_x$
Metallic region
$j_y^A$
$j$
$\theta^A$
$j_x$
$\tan\theta_1^A$
$f^1$
$f^3$
$f^5$
$j_y^A << j_x$
$\tan\theta^A$
$\rho_{xx}\sigma_H^A$
b
$\tan\theta^A$
$\rho_{xx}$ (μΩ cm)
$\sigma_H^A$ ($10^3$ $\Omega^{-1}$ cm$^{-1}$)
x
y
z
c
Transition region
Intrinsic region
2.0
1.5
1.0
0.5 = $\sigma_H^{in}$ ($10^3$ $\Omega^{-1}$ cm$^{-1}$)
$\tan\theta^A$
$\rho_{xx}$ (μΩ cm)

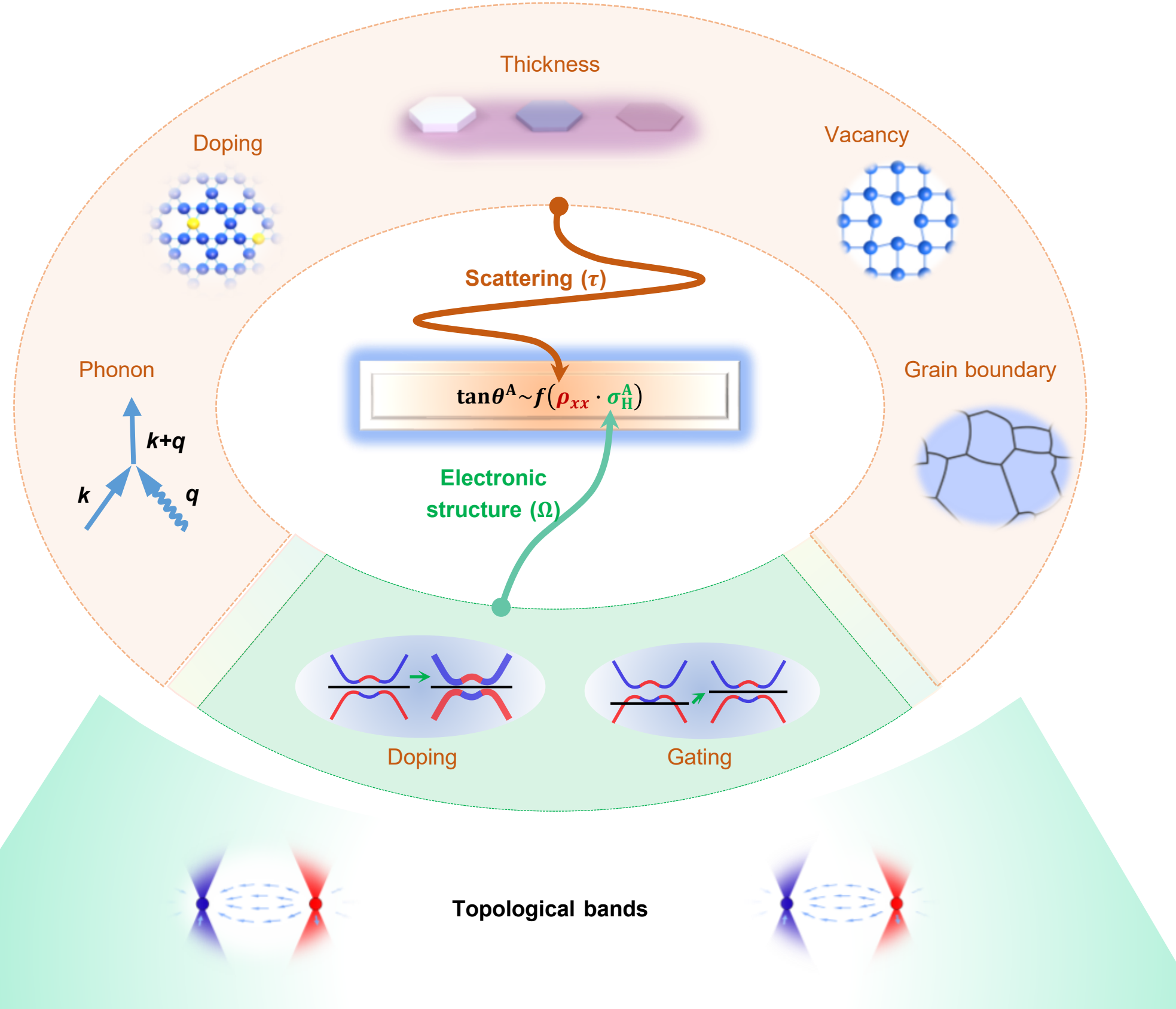
Thickness
Doping
Vacancy
Scattering ($\tau$)
Phonon
Grain boundary
$\tan\theta^{A} \sim f(\rho_{xx} \cdot \sigma_{H}^{A})$
k+q
k
q
Electronic structure ($\Omega$)
Doping
Gating
Topological bands

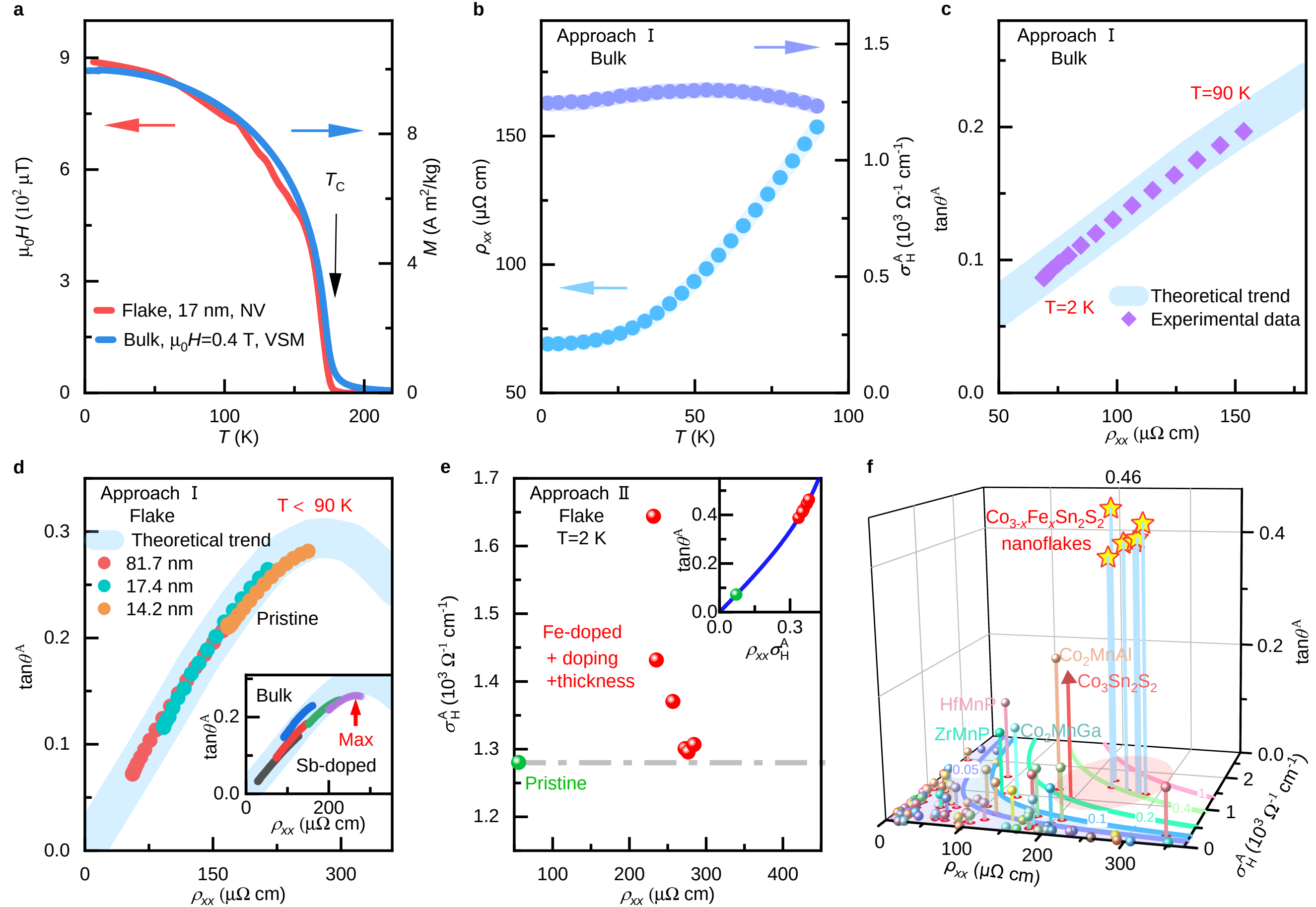
a
Flake, 17 nm, NV
Bulk, μ0H=0.4 T, VSM
T_C
μ0H (10^2 μT)
M (A m^2/kg)
T (K)
b
Approach Ⅰ
Bulk
ρxx (μΩ cm)
σ_H^A (10^3 Ω^-1 cm^-1)
T (K)
c
Approach Ⅰ
Bulk
T=90 K
T=2 K
Theoretical trend
Experimental data
tanθ^A
ρxx (μΩ cm)
d
Approach Ⅰ
Flake
T < 90 K
Theoretical trend
81.7 nm
17.4 nm
14.2 nm
Pristine
Bulk
Max
Sb-doped
tanθ^A
ρxx (μΩ cm)
e
Approach Ⅱ
Flake
T=2 K
Fe-doped
+ doping
+thickness
Pristine
σ_H^A (10^3 Ω^-1 cm^-1)
ρxx σ_H^A
ρxx (μΩ cm)
f
0.46
Co3-xFexSn2S2
nanoflakes
Co2MnAl
Co3Sn2S2
HfMnP
ZrMnP
Co2MnGa
ρxx (μΩ cm)
σ_H^A (10^3 Ω^-1 cm^-1)
tanθ^A

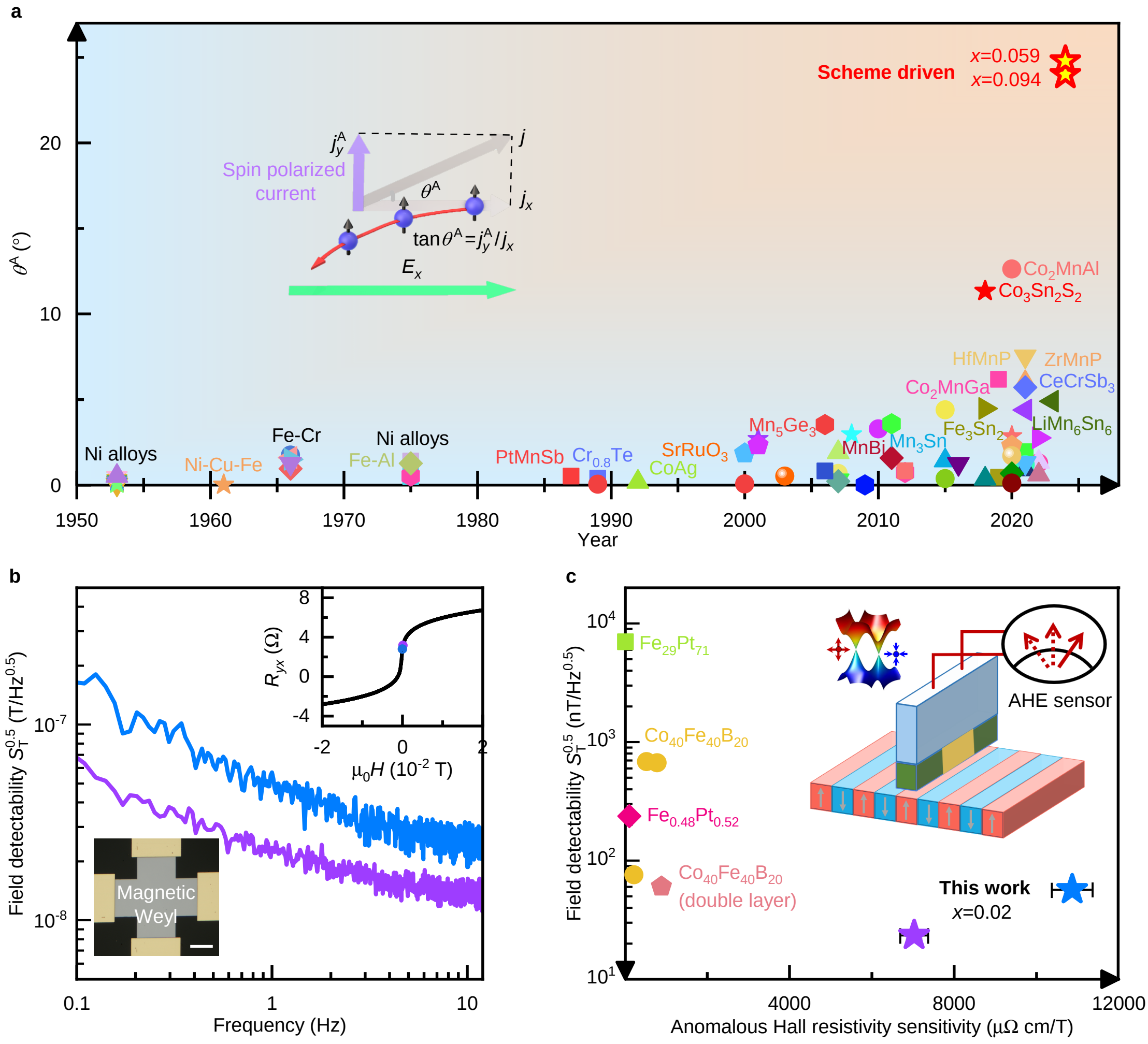
a
Scheme driven
x=0.059
x=0.094
Spin polarized current
tan θA=jyA/jx
Ex
Co2MnAl
Co3Sn2S2
HfMnP
ZrMnP
Co2MnGa
CeCrSb3
LiMn6Sn6
Fe3Sn2
Mn3Sn
MnBi
Mn5Ge3
SrRuO3
CoAg
PtMnSb
Cr0.8Te
Ni alloys
Fe-Al
Fe-Cr
Ni-Cu-Fe
Year
θA (°)
b
Field detectability ST0.5 (T/Hz0.5)
Frequency (Hz)
Ryx (Ω)
μ0H (10-2 T)
Magnetic Weyl
c
Fe29Pt71
Co40Fe40B20
Fe0.48Pt0.52
Co40Fe40B20 (double layer)
AHE sensor
This work
x=0.02
Field detectability ST0.5 (nT/Hz0.5)
Anomalous Hall resistivity sensitivity (μΩ cm/T)